\documentclass[ 
preprint,
showpacs,
preprintnumbers,
amsmath,amssymb,
aip,
floatfix,
]{revtex4-1}

\usepackage{graphicx}% Include figure files
\usepackage{bm}% bold math

\begin{document}

\title{Comment on "Ion velocity analysis of rotating structures in a magnetic linear plasma device" [Phys. Plasmas 25, 061203 (2018)] }

\author{Thi\'{e}ry PIERRE}

\affiliation{Centre National de la Recherche Scientifique, 31 Chemin Joseph Aiguier, Marseille, France.}

%\date{\today}

\begin{abstract}
In a recent paper (Phys. Plasmas 25, 061203, 2018), the authors have presented the analysis of the electric ion drift velocity experienced by heavy ions created in a plasma submitted to a low magnetic field. Unfortunately, they have used the classical ExB drift formula that is valid only in slab geometry. The authors have not taken into account that the cylindrical geometry induces a slow electric drift of the ions around the axis of the column. Moreover, the low magnetization of the ions induces a Larmor radius that is larger than the diameter of the plasma column. The movement of the ions immediately after their creation is parallel to the local electric field, not perpendicular as indicated by the authors. Most often the ions are neutralized before experiencing the electric drift calculated along the classical guiding center theory. This has not been taken into account carefully by the authors so that the theoretical analysis of the Laser Induced Fluorescence measurements presented in this paper is clearly invalid.

\end{abstract}


\maketitle

In the paper "Ion velocity analysis of rotating structures in a magnetized linear plasma device" \cite{claire} , the authors report the use of the Laser Induced Fluorescence (LIF) technique in order to analyze the movement of the ions at the periphery of a magnetized plasma when nonlinear electrostatic structures rotate around the plasma.

Rotating structures in magnetized plasmas have been early observed in laboratory plasmas. In fact, the rotation of a magnetized plasma column submitted to an inward electric field has been observed more than half a century ago \cite{vlasov}. Spiral structures were detected then. The spiralling plasma at the edge of a magnetized plasma column has been reported later by Ikehata et al. \cite{ikehata}. In the MISTRAL device, we have observed similar structures fifteen years ago \cite{matsuk, barni, pierre} and we have carefully analyzed our observations. A visualization technique has been then applied using an intensified gated camera and we have demonstrated the presence of plasma structures rotating at the periphery of the magnetized plasma column of the MISTRAL device \cite{jaegerPoP}. Surprisingly, our measurements (Fig. 2 in the commented paper) originating from a document produced 15 years ago \cite{jaegerPhD} are displayed as original results obtained by the authors. During the 2010 measurement campaign, the LIF technique improved the analysis of the plasma structures around the magnetized plasma column. The results have been detailed in a PhD manuscript (2010) \cite{rebont}. They are now reported very late by N. Claire et al. in the recent paper commented here.

 Clearly, the LIF technique allows to obtain a precise measurement of the velocity distribution function of the argon ions. In order to derive the spatiotemporal structure of the electric field from these measurements, the authors refer to a model relating directly the ion velocity to the electric field using the simple E/B formula. This model is highly questionable. In fact, it is important to distinguish two different situations.
 
 The first situation is obtained when very slow ions are taken into account. The ion Larmor radius is then small compared to the radius of the plasma column. This has been first analyzed by A. B. Mikhailovskii \cite{mikhailov, mikhailov1} more than half a century ago, assuming a rigid body rotation of the plasma column. In this situation, the guiding center theory in cylindrical geometry can be used. The drift of the ions (more precisely here the angular frequency $\Omega_z$ of the ion cloud) considering an inward electric field, with the electric drift frequency $\Omega_E = E/ (r.B)$ and the ion cyclotron frequency $\Omega_{ci}$ , is obtained following the expression where a rigid body rotation is assumed:
 
 \begin{align}
	\label{eq:slow_velocity}
	 \qquad \Omega_z &= \dfrac{- \Omega_{ci}}{2} 
	\left( 1 - \sqrt{1+4\dfrac{\Omega_E}{\Omega_{ci}}} \right) \\
\end{align}

It is important to note that the classical formula Vdrift= E/B (slab model) is valid in cylindrical geometry only considering a very low electric field as well as a large radius of the plasma column. This is never the case in the experiment under consideration in the commented paper. In fact, the cyclotron movement of the ions leads to an azimuthal drift along a circle centered on the axis of the plasma column. The above expression is simply due to the cylindrical geometry. Fifteen years ago, we have extended the theoretical analysis to the specific case of our laboratory plasma column calculating the angular rotation frequency of both electrons and ions around the axis of the plasma column. We have estimated the growth rate of the instability \cite{sosenko}. The transition from an azimuthal mode m=1 to a mode m=2 has been predicted and compared to the experiment.

In the second situation, ions are moving with a velocity of the order of the ion thermal velocity and larger. It is important to bear in mind that fast ions are created as a result of wave-particle interaction inside the moving nonlinear electrostatic structure. For instance, ions moving at thermal velocity experience a radial excursion comparable to the radius of the plasma column. Faster ions are not at all confined by the magnetic field. Moreover the gas pressure used in the experiment leads to a high probability for ions to be neutralized by charge exchange collision after travelling a few centimeters.

The second point to consider regarding the results presented by the authors is the following. They mention a different result of the measurements when they consider an azimuthal mode m = 1 or a mode m = 2. They write that this is difficult to understand and that it would require numerical simulations.
  
The authors should take into consideration that it is perfectly clear that the two cases are quite different concerning radial ion velocities.
In the case of the mode m = 1, as we showed fifteen years ago on this same experimental device, the m = 1 mode corresponds to a global eccentric movement of the plasma column (see for instance Fig. 6 in our paper published fourteen years ago \cite{barni}). As a result, a mean radial ion velocity is present at the edge of the plasma column. On the other hand, the mode m = 2 corresponds to an axisymmetric movement of the column with two arms presenting spiral extensions around the column. It is quite clear in this case that the ion motion is preferentially azimuthal with lower radial velocity.

 In summary, we have presented reasons for rejecting the interpretation of the results published by N. Claire et al. using Laser Induced Fluorescence when measuring the ion velocity distribution at the edge of a rotating magnetized plasma column.

\end{document}